\documentstyle[12pt]{article}
\begin{document}
\begin{center}   {\Large \bf PION MASS EFFECTS IN THE LARGE $N$ LIMIT OF $\chi
PT$}
 \vskip 1cm
{\large   \bf Antonio  DOBADO} \footnote{E-mail:
dobado@cernvm.cern.ch}    \\ {\large  Departamento de F\'{\i}sica Te\'orica \\
Universidad Complutense de Madrid,
 28040 Madrid, Spain  \\ and \\} \vskip .5 cm {\large \bf John MORALES}
\footnote{On leave of absence from  Centro Internacional de F\'{\i}sica,
Colombia. }  \footnote{E-mail: jmorales@ccuam3.sdi.uam.es }  \\ {\large
Departamento de F\'{\i}sica Te\'orica \\ Universidad Aut\'onoma de Madrid,
28049 Madrid, Spain} \vskip 1cm  July 1994  \vskip 1cm

\begin{abstract}

We compute the  large $N$ effective action of the $O(N+1)/O(N)$  non-linear
sigma model  including the effect of the pion mass
 to order $m^2_{\pi}/f_{\pi}^2$. This
 action is more complex than the one corresponding to the chiral limit not
only because of the pion propagators but also because chiral symmetry produce
new interactions
 proportional to $m^2_{\pi}/f_{\pi}^2$. We renormalize the action by including
the appropriate  counter terms and find the renormalization group equations
for the corresponding
 couplings. Then we estudy the unitarity propierties of the scattering
amplitudes. Finally our results are applied to the particular case of the
linear sigma model and also are used to fit the pion scattering  phase shifts.

\end{abstract}
\end{center}
FT/UCM/18/94,
\newpage
\baselineskip 0.83 true cm \textheight 20 true cm

\section{Introduction}

In recent years a lot of effort has been devoted to the so called
 Chiral Perturbation Theory $(\chi PT$) \cite{Wei,GassLeut}. This  formalism
provides an usefull tool for the phenomenological description of the
low-energy hadron dynamics in terms of some parameters that can be fitted from
the experimental data. Even it is possible that some day they will be computed
from the underlying theory which is expected to be Quantum Cromodynamics
(QCD).

As $\chi PT$ can be applied without a precise formulation of this underlying
theory, it has been also used for the parametrization of the unknown  Symmetry
Breaking Sector of the
 Standard Model \cite{DoHe} giving rise to a model independent description of
this sector that can be quite usefull for the analysis of the future Large
Hadron Collider
 (LHC) data.

However, $\chi $PT still has some problems that should be solved in order to be
completely usefull for the practical applications. Usually  $\chi $PT
 computations are done to the one-loop  level, or what it is the same,
expanding the amplitudes to the forth power of the momenta or pion masses (see
\cite{2loop} for an exception to this rule). Thus, the region of
 applicability of  $\chi $PT is restricted to  the very low-energy regime.
 In fact, $\chi $PT does not satisfy exactly the unitarity conditions but only
in the perturbative sense, and this problem becomes very relevant at higher
energies \cite{Truong}. To avoid this limitation of $\chi PT$ several methods
have been proposed like the introduction  of new fields corresponding to
resonances \cite{Resonances}, the Pad\'e approximants \cite{Truong}, the
inverse amplitude method  \cite{Truong,piK}, etc. More recently it has also
been considered the so called large $N$ expansion of $\chi $PT ($N$ being the
number of Nambu-Goldstone bosons). This approximation is not restricted to the
low energy domain and from the point of view of unitarity it has very good
behavior. Until now this approach has been used for the computation of the
pion scattering amplitudes \cite{largen,DoLoMo} in the chiral limit and also
it was used in \cite{Im} to study the $\gamma \gamma
 \rightarrow \pi^0 \pi^0$ reaction. Some of the results in \cite{largen}
 were also  reobtained in \cite{Dugan} but using  different techniques. The
work in \cite{Im} has been, we think properly,  criticised in \cite{Penn}, but
work is in progress in order to make a correct computation of the $\gamma
\gamma
 \rightarrow \pi^0 \pi^0$ reaction in the large $N$ limit.

The main goal of this work is to extend the work in \cite{largen,DoLoMo}
outside of the chiral limit i.e. we study the pion effective action and the
pion scattering in the large $N$ limit taken into account the effect of the
pion mass. As it is well known this is not trivial at all since, because of
the chiral symmetry, the introduction of the pion mass  give rise to new
interactions.

The plain of the paper goes as follows: In section two we introduce the
dynamics of the $O(N+1)/O(N)$ Non-Linear Sigma Model (NLSM) and we compute the
effective action up to order $1/N$ including the leading corrections coming
from having a pion mass different from zero. In section three we obtain the
renormalized elastic pion scattering amplitude in terms of the renormalized
coupling constants. In section four we study the running of those couplings
and the renormalization of the pion mass and the pion decay constant. In
section five we discuss the unitarity properties of the amplitudes. Section
six is devoted to the particular case of the Linear Sigma Model (LSM) which is
a very nice example to test the previously developped formalism. In section
seven we show a fit of the experimental data for the $I=J=0$ elastic scattering
 amplitude and finally the conclussions are presented in section eight.

\section{The dynamics of the $O(N+1)/O(N)=S^N$ non-linear sigma model}

 In order to define  the large $N$ limit of $\chi $PT we start from the
two-flavor chiral symmetry group $SU(2)_R \times SU(2)_L$, and then the
equivalence of the coset spaces $SU(2)_L \times
SU(2)_R/SU(2)_{L+R}=O(4)/O(3)=S^3$ is used to extend this symmetry pattern to
$O(N+1)/O(N)=S^N$. Therefore $N$ is the dimension of the coset space  and the
number of Nambu-Goldstone bosons (NGB), or in other words, the pions. The NGB
fields can be chosen as arbitrary coordinates on $S^N$. The most general
$O(N+1)$ invariant lagrangian can be written as
 a derivative expansion which is covariant with respect to both the space-time
and
 the $S^N$ coordinates. The lowest order is given by:
\begin{equation} {\cal
L}_{0}=\frac{1}{2}g_{ab}(\pi) \partial _{\mu}\pi^a\partial^{\mu}\pi^b
 \end{equation}
When using the standard coordinates the metrics is found to be:
\begin{eqnarray} g_{ab}= {\delta}_{ab} +\frac{1}{NF^2} \frac{{\pi_a}{\pi_b}}
{1-\frac{\pi^2}{NF^2}}
\end{eqnarray}
 which has been obtained from the free lagrangian for the $N+1$ fields $\pi_1,
\pi_2,...\pi_N,\sigma$ by forcing them to live in the sphere $\pi^2+\sigma^2=
\Sigma_{a=1}^N\pi_a\pi_a+\sigma^2=NF^2$. As it is well known this lagrangian
not only contains the kinetic term but also an infinite number of interacting
terms
 with an arbitrary even number of pions. If we now want to introduce the pion
mass we have to explicitly break the $O(N+1)$ symmetry while keeping the $O(N)$
one. This can be achieved just considering in the lagrangian a properly
normalized term proportional to the $\sigma$ component. Then the total
lagrangian becomes:

\begin{eqnarray}
{\cal L}= \frac{1}{2} g_{ab} {\partial}_{\mu} {\pi^a}
{\partial}^{\mu} {\pi^b} +NF^2m^2 \sqrt{1-\frac{\pi^2}{NF^2}} \end{eqnarray}
where $m$ plays the role of the pion mass as it can be seen by expanding the
squared root. In addition we will also find another infinite set of terms
which produce new pion interactions.

Therefore the lagrangian in eq.3 will be our starting point to study the pion
dynamics in the large $N$ limit. As usual the action is given by:
\begin{eqnarray} {\cal S} \left[\pi\right]= \int d^nx {\cal L}
\left({\pi}\right) \end{eqnarray}

However, to define the quantum theory we need to know not only the action but
also the  measure in the field space which defines the path integral
representation of the Green functions. For the case of the NLSM, the
appropriate
 measure includes a factor which is the squared root of the coset metrics
determinant \cite{Charap}. This factor contributes to the lagrangian with a
term proportional to $\delta^n(0)$. However, it is well known that using
dimensional regularization \cite{Tararu}, where  $n=4-\epsilon$, it is valid
the following rule, $\delta ^n(0)=0$ or equivalently $\int d^nk=0$, and
therefore the measure term can be ignored (for a recent discussion about
regularization methods in $\chi$PT see \cite{Espriu}).

Thus the generating functional for the regularized Green functions will be
given by:
\begin{eqnarray} e^{iW\left[J\right]}= \int \left[d\pi\right] e^{i
\left( {\cal S} \left[ \pi \right] +\int d^nx J^a{\pi^a} \right) }
\end{eqnarray} with $n=4-\epsilon$.

As it was discussed in the introduction we are interested in the  computation
of the effective action up to order $1/N$. Therefore we need some  systematic
method to obtain the diagrams that contribute in this approximation. This can
be done using a generalization of the technique discussed in \cite{Coleman}
for the case of the linear model, which  introduces two auxiliary fields
$B_{\mu}$ and $\phi$. The first one is connected with the derivative
interactions whereas the second is related to those interactions proportional
to $m^2$. The lagrangian including these auxiliary fields is defined as:

\begin{eqnarray} \tilde{\cal L} & = & {\cal L} -\frac{1}{2}NF^2 \left( B_{\mu}
-\frac{{\pi^a}{\partial}_{\mu}{\pi^a}} {NF^2\sqrt{1-\frac{\pi^2}{NF^2}}}
\right)^2 \nonumber\\ & + & \frac{1}{2}NF^2m^2 \left( \phi- \sqrt{2} \left(1
-\frac{\pi^2}{2NF^2} -\sqrt{1-\frac{\pi^2}{NF^2}} \right) \right)^2
\end{eqnarray}

whose main properties are the following: first of all, since there is no
kinetical term for  the new auxiliary fields, they can be integrated out in an
easy way,  and then one finds:

\begin{eqnarray} e^{i{\cal S}\left[\pi\right]}\approx \int
\left[dB\right]\left[d\phi\right] e^{i \int d^nx \tilde{\cal L}
\left(\pi,B,\phi\right) } \end{eqnarray}

Second, the new self-interaction terms appearing in $\tilde{\cal L}$ have
been  chosen so that they cancel the old ones appearing in ${\cal L}$ and thus
the new action can be written as follows:

\begin{eqnarray} \tilde{\cal S}\left[\pi,B,\phi\right] & = & \int d^nx \left\{
NF^2m^2 -\frac{1}{2}{\pi^a} \left(\Box+m^2\right) {\pi^a}\right. \nonumber\\ &
- & \left. \frac{1}{2}NF^2 \left( B^2-m^2{\phi^2} \right)
-\frac{1}{2}NF^2{\partial}_{\mu}B^{\mu} f\left(\frac{\pi^2}{NF^2}\right)\right.
\nonumber\\ & - & \left. \sqrt{2}NF^2m^2{\phi} g\left(\frac{\pi^2}{NF^2}\right)
\right\} \end{eqnarray} where: \begin{eqnarray}
f\left(\eta\right)&=&2\left(1-\sqrt{1-\eta}\right) \nonumber\\
g\left(\eta\right)&=&\left(1-\frac{\eta}{2}-\sqrt{1-\eta}\right)^ {\frac{1}{2}}
\end{eqnarray} This action describes the same dynamics for the pions than that
in eq.3 and eq.4 but
 now, instead of self-interactions, we have $B_{\mu}$ and $\phi$ mediated
 interactions. This fact makes it easier to trace out the $N$ factors for  a
given diagram, and therefore to select those corresponding to the different
terms in the Green functions $1/N$ expansion.

{}From the action in eq.8  it is an straightforward matter to obtain the
Feynman
rules. To select the relevant diagrams for the leading order in the $1/N$
expansion the most important thing is to know the $N$ power of the different
propagators and vertex involved. It is immediate to check that the $B$ and
$\phi$ propagators are just constants proportional to $1/N$, the $B2\pi$ and
$\phi 2\pi$ vertices are of the order one,  the $B4\pi$ and the  $\phi 4\pi$
vertices are of the order $1/N$, the $B6\pi$ and the  $\phi 6\pi$ vertices are
of the order $1/N^2$ and so on.

The most general one-particle-irreducible ($1PI$) pion Green function can be
obtained by attaching pions to the leg of the most general $1PI$ $B$ and
$\phi$ Green function (see fig.1). Those $1PI$ $B$ and $\phi$ Green functions
(represented by a black dot) can be expanded in terms of other reduced
 $B$ and $\phi$ $1PI$ functions (represented by a dashed dot) which are defined
as the sum of all the diagrams containing only pion loops but not $B$ or
$\phi$ loops  (see fig.2 for an example). The reduced $B$ and $\phi$ Green
functions are as much of   order $N$ (see fig.3). In addition, the diagrams
with $B$ or $\phi$ loops contributing to the general $B$ and $\phi$ Green
functions are suppressed in the $1/N$ expansion (because of the $1/N$ factors
in the $B$ and $\phi$
 propagators) when compared with the first term. Thus we find the symbolic
equation shown in fig.4. As usual, the connected Green functions can be
expressed in terms of the $1PI$ ones by making trees, so that only the latter
will be considered in detail.

The $B$ and $\phi$ $1PI$ (dashed) Green functions can be obtained  from the
effective action: \begin{eqnarray} e^{i\Gamma\left[B,\phi\right]}= \int
\left[d\pi\right] e^{i \int d^nx \tilde{\cal L} \left(\pi,B,\phi\right) }
\end{eqnarray} In fig.5 we show the diagrams contributing at leading order in
$1/N$ to those $1PI$ Green functions with just one or two $B$ and $\phi$ legs.
It is not difficult to see that diagrams with three or more $B$ or $\phi$ legs
do not contribute at the level of precision considered in this paper, i.e. to
the
 leading order in the $1/N$ expansion and to the order $m^2/F^2$.

Now, using standard functional technics, it is possible to find:

\begin{eqnarray} \Gamma\left[B,\phi\right] & = &  \int d^nx \left\{ NF^2m^2-
\frac{1}{2}NF^2 \left(B^2-m^2{\phi^2}\right)\right \} \nonumber\\ & - &
\sqrt{2}NF^2m^2 g\left(\frac{I_m}{F^2}\right) \int d^nx\phi\left(x\right)
\nonumber\\ & - & 2Nm^4 g'^2 \left( \frac{I_m}{F^2} \right) \int d^nxd^ny
\phi\left(x\right) K\left(x-y\right) {\phi\left(y\right)} \nonumber\\ & - &
\frac{N}{4} f'^2 \left( \frac{I_m}{F^2} \right) \int d^nxd^ny
B_{\mu}\left(x\right) G^{\mu\nu}\left(x-y\right) B_{\nu}\left(y \right)
\nonumber\\ & - & i\sqrt{2}Nm^2 f'\left(\frac{I_m}{F^2}\right)
g'\left(\frac{I_m}{F^2}\right) \int d^nxd^ny B_{\mu}\left(x\right)
G^{\mu}\left(x-y\right) {\phi}\left(y\right) \nonumber\\ & + & ...
\end{eqnarray}

where the points mean terms of $1/N^2$ order as well as terms with more than
two $B$ or $\phi$ fields. Notice that the above effective action is not local
being the integral kernels defined by:

\begin{eqnarray} K\left(x-y\right) & = & \int d\tilde{k}
e^{-ik\left(x-y\right)} I\left(k^2\right)  \\     \nonumber
G_{\mu\nu}\left(x-y\right) & = & \int d\tilde{k} e^{-ik\left(x-y\right)}
k_{\mu}k_{\nu} I\left(k^2\right)  \\       \nonumber G_{\mu}\left(x-y\right) &
= & \int d\tilde{k} e^{-ik\left(x-y\right)} k_{\mu} I\left(k^2\right)
\end{eqnarray}

where for short we have defined

\begin{eqnarray} d\tilde{k}\equiv {\mu^{\epsilon}} \int
\frac{d^{4-{\epsilon}}k} {\left(2{\pi}\right)^{4-{\epsilon}}} \end{eqnarray}

and the loop integrals $I_m$ and $I(k^2)$ are given by:

\begin{eqnarray}I_{m}= \int d\tilde{q} \frac{i}{q^2-m^2}\equiv
-m^2\bigtriangleup   \\ \nonumber I\left(k^2\right)= \int d\tilde{q} \frac{i}
{\left[\left(k+q\right)^2-m^2\right] \left(q^2-m^2\right)} \end{eqnarray}

Using dimensional regularization these integrals are found to be:

\begin{eqnarray} I_{m}= \frac{-m^2}{\left(4{\pi}\right)^2} \left\{
N_{\epsilon}+1-\log{\frac{m^2}{\mu^2}} \right\} \end{eqnarray}

and

\begin{eqnarray} I\left(k^2\right)= \frac{-1}{\left(4{\pi}\right)^2} \left\{
N_{\epsilon}+2+ \sqrt{1-\frac{4m^2}{k^2}} \log{ \frac{
\sqrt{1-\frac{4m^2}{k^2}}-1 } { \sqrt{1-\frac{4m^2}{k^2}}+1 } }
-\log{\frac{m^2}{\mu^2}} \right\} \end{eqnarray}

where, as usual:

\begin{eqnarray} N_{\epsilon}=\frac{2}{\epsilon}+\log{4\pi}-\gamma
\end{eqnarray}

Now we couple the effective action for the $B$ and $\Phi$ fields to the pions
by defining the action:

\begin{eqnarray} \tilde{\cal S}\left[\pi,B,\phi\right] & = &  \int d^nx \left\{
\frac{-1}{2} {\pi^a} \left(\Box +m^2\right) {\pi^a}
-\frac{1}{2}NF^2{\partial}_{\mu}B^{\mu} f\left(\frac{\pi^2}{NF^2}\right)\right.
\nonumber\\ & - & \left.\sqrt{2}NF^2m^2{\phi} g\left(\frac{\pi^2}{NF^2}\right)
\right\} +\Gamma \left[B,\phi\right] \end{eqnarray} As discussed before, the
pion Green functions can be obtained by attaching pions to the $B$ and $\Phi$
Green functions. However it
 should be noted that it is also possible to have pion loops such as the one
shown in fig.1, which are proportional to the integral $I_m$ and therefore, to
$m^2$. Thus, from the formal point of view, the pion effective action can be
written as:

\begin{eqnarray} e^{i {\cal S}_{eff} \left[\pi\right] }= \int
\left[d{\pi'}\right] \left[dB\right] \left[d{\phi}\right] e^{i \tilde{\cal S}
\left[\pi',B,{\phi}\right] } \end{eqnarray}

where the integral on the $B$ and $\Phi$ fields must  be done at  tree level.
Moreover, since we want to expand the pion effective action
 only up to order $m^2/F^2$ terms, we only have to perform the integral on the
pion fields to the one loop level. Therefore, to the  considered degree of
accuracy, we have:

\begin{eqnarray} \tilde{\cal S}\left[{\pi'},B,\phi\right] & = &  {\cal S}
\left[{\pi'},\bar{B},\bar{\phi}\right] \nonumber\\ & + & \frac{1}{2} \int
d^nxd^ny \frac{\delta\tilde{\cal S}}
{\delta\pi^a\left(x\right)\delta\pi^b\left(y\right)} |_{{\pi'}=\pi}
{\pi'}_{a}\left(x\right) {\pi'}_{b}\left(y\right)+.... \end{eqnarray}

where $\bar{B}$, $\bar{\phi}$ and $\pi$ are defined so that:

\begin{eqnarray} \frac{\delta\tilde{\cal S}} {\delta B_{\mu}
\left(x\right)}=0    \\ \nonumber \frac{\delta\tilde{\cal S}} {\delta
\phi\left(x\right)}=0
  \\ \nonumber \frac{\delta\tilde{\cal S}} {\delta{\pi^a}\left(x\right)}=0
\end{eqnarray}
 for $ B=\bar{B}, \phi=\bar{\phi}$ and $\pi'={\pi}$.  These equations can be
used to write  $\bar{B}$ and $\bar{\phi}$ in terms of $\pi$ as follows:

\begin{eqnarray} \bar{B}_{\mu}\left(x\right) & = &  -\int d^nyd\tilde{k}
e^{-ik\left(x-y\right)} \frac{ik_{\mu}} {1+\frac{k^2I\left(k\right)}{2F^2}}
\nonumber\\ & \times & \frac{\pi^2\left(y\right)}{2NF^2} \left\{1+
\frac{I\left(k\right)m^2} {2\left[1+ \frac{k^2I\left(k\right)}{2F^2} \right]F^2
} \left(1+\bigtriangleup\frac{k^2}{F^2}\right) \right\}+... \end{eqnarray}

and

\begin{eqnarray} \bar{\phi}\left(x\right) & = &  \frac{1}{2}\int d^nyd\tilde{k}
e^{-ik\left(x-y\right)} \frac{1} {1+\frac{k^2I\left(k\right)}{2F^2}}
\nonumber\\ & \times & \left\{ \frac{\pi^2\left(y\right)}{NF^2} +\frac{1}{4}
\left( \frac{\pi^2\left(y\right)}{NF^2} \right)^2 \left(
1-\frac{k^2I\left(k\right)}{2F^2} \right) \right\}+... \end{eqnarray}

If we now integrate the $B$ and $\Phi$ fields to the tree level and the pion
field to the one-loop level we can write:

\begin{eqnarray} {\cal S}_{eff}\left[\pi\right]= \tilde{\cal
S}\left[\pi,\bar{B},\bar{\phi}\right] +\frac{i}{2} Tr\log{ \frac{\delta^2{\cal
S}} {\delta\pi^a\left(x\right)\delta\pi^b\left(y\right)} }+... \end{eqnarray}

where:

\begin{eqnarray} \frac{\delta^2{\cal S}} {\delta\pi^a(x)\delta\pi^b(y)} & = &
-\left(\Box +m^2\right) {\delta}_{ab} \delta(x-y)
+2{\partial}_{\mu}B^{\mu}f'\left(\frac{I_m}{F^2}\right)
\frac{\pi^a(x)\pi^b(y)}{NF^2} \nonumber\\ & + &
{\partial}_{\mu}B^{\mu}f'\left(\frac{I_m}{F^2}\right) {\delta}_{ab}
\delta\left(x-y\right)+... \nonumber\\ & = & D_{ab}(x,y)+\triangle_{ab}(x,y)
\end{eqnarray}

with

\begin{equation}
 D_{ab}(x,y)=-\left(\Box +m^2\right){\delta}_{ab}\delta(x-y) \end{equation}

and $\triangle_{ab}(x,y)$ can be obtained from the above two equations. Now it
is possible to write:

\begin{eqnarray} Tr \log{\left(D+\triangle\right)} & = &  Tr
\log{D\left(1+D^{-1}\triangle\right)} \nonumber\\ & = & Tr
\log{D}+\sum_{n=1}^{\infty} \frac{\left(-1\right)^{n+1}}{n}
\left(D^{-1}\triangle\right)^n \end{eqnarray}

In the following we will ignore the $Tr \log{D}$ term since it does not depend
on the pion fields and therefore its contribution to the  pion effective
action is irrelevant. On the other hand it is not difficult to see that only
the term $n=1$ contributes to the effective action  up to the level of the
approximations considered here. Then, using the above equations it is possible
to arrive to:

\begin{eqnarray} {\cal S}_{eff}\left[\pi\right] & = &  \int d^4x
\frac{-1}{2}{\pi^a} \left\{\Box +m^2 \left(1+\frac{I_m}{2F^2}\right)
\right\}{\pi^a} \nonumber\\ & + & \frac{1}{8NF^2} \int d^4xd^4y \int \frac{d^4
q} { \left(2\pi\right)^4 } e^{-iq\left(x-y\right)} {\pi^2}\left(x\right)
{\pi^2}\left(y\right) \nonumber\\ & \times &
\frac{1}{1+\frac{q^2I\left(q\right)}{2F^2}} \left\{ q^2-
\frac{m^2\left(1+\bigtriangleup\frac{q^2}{F^2}\right)}
{1+\frac{q^2I\left(q\right)}{2F^2}} \right\} \nonumber\\ & + &
O\left[\left(\frac{m^2}{F^2}\right)^2\right]+ O\left[\frac{1}{N^2}\right]
\end{eqnarray} which is our final result for the regularized pion effective
action to leading order in the $1/N$ expansion and to order $m^2/F^2$.
\section{The Scattering Amplitude}
%
%
%
%
immediate to find the scattering amplitude  for the process $\pi_a\pi_b
\rightarrow \pi_c\pi_d$ from the pion effective action obtained in the
preceding section. This amplitude can be written as: \begin{eqnarray} T_{abcd}=
\delta_{ab}\delta_{cd}A\left(s\right)+ \delta_{ac}\delta_{bd}A\left(t\right)+
\delta_{ad}\delta_{bc}A\left(u\right) \end{eqnarray}

where:

\begin{eqnarray} A\left(s\right)= \frac{1}{NF^2} \left\{ \frac{s}
{1+\frac{sI\left(s\right)}{2F^2}} -\frac{m^2} {\left(1+
\frac{sI\left(s\right)}{2F^2} \right)^2} \left(1+\bigtriangleup
\frac{s}{F^2}\right) \right\} \end{eqnarray}

It is not difficult to see that the above amplitude corresponds to the
diagrams appearing in fig.6, which are those relevant to the leading order in
the $1/N$ expansion and to order $m^2/F^2$. However, for practical
applications, this regularized amplitude should be  renormalized. Moreover,
the renormalization procedure is not trivial
 since the NLSM is not renormalizable in the standard sense. As it was
 discussed in \cite{largen} the renormalization of the NLSM in the large $N$
limit requires the introduction of an infinite number of counter terms in the
renormalized lagrangian. In the case considered here things are even  harder
since, in addition to the derivative interactions that appear in the  chiral
limit, we also have new interactions proportional to $m^2$.

Therefore, we will introduce the renormalized lagrangian:

\begin{eqnarray} {\cal L}_{R}={\cal L}+c.t \end{eqnarray}

where $c.t.$ represents the infinite set of counterterms needed to absorb all
the divergences appearing in the the effective action in eq.28.  In fact, at
the
 level considered here, we do not need to know the precise form of these
counterterms. From fig.6  we know that the only vertices appearing in the
diagrams contributing to the regularized pion scattering amplitude are the
derivative four pion vertex, the four pion vertex proportional to $m^2$ and
the derivative six pion vertex. The effect of the counterterms needed for the
renormalization  of the pion scattering amplitude on these vertices can be
easily taken into account just making the following replacements:

\begin{eqnarray} \frac{q^2}{NF^2}&\longrightarrow&\frac{q^2}{NF^2}G
\end{eqnarray}

on the derivative four pion vertex

\begin{eqnarray} -\frac{m^2}{NF^2}&\longrightarrow&-\frac{m^2}{NF^2}H
\end{eqnarray}

on the four pion vertex proportional to $m^2$ and

\begin{eqnarray} \frac{2q^2}{\left(NF^2\right)^2}&\longrightarrow&
\frac{2q^2}{\left(NF^2\right)^2}E \end{eqnarray}

on the derivative six pion vertex (in the above prescriptions $q$  represents
the total momentum of two of the pions involved in the vertex although for the
sake of simplicity we do not give the details of the appropriate combinatory
). The factors $G$, $H$ and $E$ are arbitrary analytical functions on $s$ so
that they can be written as:

\begin{eqnarray} G\left(s\right) & = & 1+g_{1}\frac{s}{F^2}+g_{2}
\left(\frac{s}{F^2}\right)^2+...
 =  \sum_{k=1}^{\infty}g_{k} \left(\frac{s}{F^2}\right)^{k} \nonumber   \\
H\left(s\right) & = & 1+h_{1}\frac{s}{F^2}+h_{2}
\left(\frac{s}{F^2}\right)^2+...
 =  \sum_{k=1}^{\infty}h_{k} \left(\frac{s}{F^2}\right)^{k} \nonumber   \\
E\left(s\right) & = & 1+e_{1}\frac{s}{F^2}+e_{2}
\left(\frac{s}{F^2}\right)^2+...
 =  \sum_{k=1}^{\infty}e_{k} \left(\frac{s}{F^2}\right)^{k} \end{eqnarray}

The  constants $g_k$, $h_k$ and $e_k$ are the unrenormalized coupling constants
and must be renormalized to absorb the divergencies in the pion scattering
amplitude. Here it is important to notice that, as it can be seen in eq.28,
the regularized action has not any new contribution to the kinetic term, and
therefore no renormalization of the pion wave function is needed. Now we can
use the above replacements to compute again the diagrams in fig.6. The new
amplitude has the same form that the previous one with

\begin{eqnarray} A\left(s\right) & = &  \frac{1}{NF^2} \left\{
\frac{sG\left(s\right)} {1+\frac{sG\left(s\right)I\left(s\right)}{2F^2}}
-\frac{m^2\left[H\left(s\right)+\bigtriangleup
E\left(s\right)\frac{s}{F^2}\right]}
{\left[1+\frac{sG\left(s\right)I\left(s\right)}{2F^2}\right]^2} \right\}
\nonumber\\ & = & A_{0}\left(s\right)+A_{1}\left(s\right) \end{eqnarray}

where $A_1(s)$ is the term proportional to $m^2$  and $A_0(s)=A(s)-A_1(s)$. As
there is no renormalization of the pion wave function, the above amplitude
should be made finite by an appropriate definition of the renormalized
couplings in terms of the unrenormalized ones. This can be done as follows;
first we will consider the renormalization of  $A_0$ and $A_1$ separately. For
the  renormalization of the $A_0$ amplitude we follow similar steps to those
followed in \cite{largen} for the renormalization of the pion scattering
amplitude in the chiral limit. Thus we write: \begin{eqnarray} A_{0}^{-1}=
\frac{NF^2}{sG}-\frac{N}{2\left(4\pi\right)^2} \left(N_{\epsilon}+2\right)-
\frac{N}{2\left(4\pi\right)^2} T\left(s;\mu\right) \end{eqnarray} where from
eq.16 and eq.17   $T(s;\mu)$ is given by \begin{eqnarray}
T\left(s;\mu\right)\equiv \sqrt{1-\frac{4m^2}{s}} \log{
\frac{\sqrt{1-\frac{4m^2}{s}}-1}{\sqrt{1-\frac{4m^2}{s}}+1} }- \log{
\frac{m^2}{\mu^2}} \end{eqnarray} Now we define $G^R$ as: \begin{eqnarray}
\frac{1}{G^{R}}=\frac{1}{G}-\frac{s}{2\left(4\pi\right)^2F^2}
\left(N_{\epsilon}+2\right) \end{eqnarray} The $G^R$ function can be expanded
in terms of the renormalized  coupling constants $g^R_k$: \begin{eqnarray}
G^{R}= 1+g_{1}^{R}\left(\frac{s}{F^2}\right)+
g_{2}^{R}\left(\frac{s}{F^2}\right)^2+... \end{eqnarray}

so that, expanding eq.39 in powers of $s/F^2$ we can find an infinite number
of equations defining the renormalized couplings $g^R_k$ in terms of the
unrenormalized ones $g_k$.

With the above definition for $G^R$, we can write: \begin{eqnarray}
A_{0}\left(s\right)= \frac{1}{NF^2} \frac{sG^{R}}
{1-\frac{sG^{R}}{2\left(4\pi\right)^2F^2}T\left(s;\mu\right)} \end{eqnarray}
In this form $A_0$ is a manifestly finite and well defined function on $s$
once  $G^R$ is given.

The next step is the renormalization of the second piece of the amplitude
$A_1$. First we note that by comparison of eq.36   and eq.41  we can write

\begin{eqnarray} \frac{G^{R}}
{1-\frac{sG^{R}}{2\left(4\pi\right)^2F^2}T(s;\mu)}=
\frac{G}{1+\frac{sGI(s)}{2F^2}} \end{eqnarray} so that $A_1$ is given by:
\begin{eqnarray} A_{1}(s)= \frac{-m^2 \left(\frac{G^R}{G}\right)^2
\left(H+\bigtriangleup E\frac{s}{F^2}\right)}
{\left(1-\frac{sG^{R}}{2F^2}T(s;\mu)\right)^2} \end{eqnarray}

Taking into account that the $A_0$ in eq.41 is finite we also have to require:

\begin{eqnarray} \frac{1}{G^2} \left[ H'- \frac{Es}{\left(4\pi\right)^2F^2}
\log{\frac{m^2}{\mu^2}} \right] \end{eqnarray}

to be finite, where:

\begin{eqnarray} H'=H+\frac{Es}{\left(4\pi\right)^2F^2}
\left(N_{\epsilon}+1\right) \end{eqnarray}

In other words, the quotients $H'/G^2$ and $E/G^2$ must be finite. Therefore
we introduce the two new generating functions $A^R$ and $B^R$ follows:
\begin{eqnarray} H' & = & A^RG^{2} \nonumber\\ E & = & B^RG^{2} \end{eqnarray}
which generate two infinite set of renormalized constants: \begin{eqnarray}
A^R & = & 1+a_{1}^R\frac{s}{F^2}+... \nonumber\\ B^R & = &
1+b_{1}^R\frac{s}{F^2}+... \end{eqnarray}

Thus, the $A(s)$ function appearing in the elastic scattering amplitude can be
written in terms of functions which are finite in the $\epsilon \rightarrow 0$
limit. \begin{eqnarray} A\left(s\right)= \frac{1}{NF^2} \left\{ \frac{sG^{R}}
{1-\frac{sG^{R}}{2\left(4\pi\right)^2F^2}T(s;\mu)} -\frac{m^2 G^{R  2}
\left[A^R-B^R\frac{s}{\left(4\pi\right)^2F^2}\log{\frac{m^2}{\mu^2}}\right]}
{\left[1-\frac{sG^{R}}{2\left(4\pi\right)^2F^2}T(s;\mu)\right]^2} \right\}
\end{eqnarray}
\section{Renormalization Group Equations}
In the last section we have been able to write the scattering amplitude in a
manifest finite form in terms of the $G^R$, $A^R$
 and $B^R$ functions. These functions generate three countable infinite sets
of renormalized couplings which implicitly depend
 on the renormalization scale $\mu$. Therefore, $G^R, A^R$ and $B^R$
 also depend on $\mu$. As we have already noticed, there is no wave function
renormalization in the leading order of the $1/N$ expansion, and thus we
conclude that the $A(s)$ amplitude is a physical observable in the sense of
the renormalization group evolution. In other words, the explicit dependence
of $A(s)$ on the renormalization scale $\mu$ must be exactly canceled by the
implicit dependence through the generating functions. This can be precisely
stated  with the evolution equation:

\begin{eqnarray} \left[\frac{\partial} {\partial{\log{\mu}}} +\sum_{k} \left(
\beta_{k}^{g} \frac{\partial} {\partial{g_{k}^R}} +\beta_{k}^{a}
\frac{\partial} {\partial{a_{k}^R}} +\beta_{k}^{b} \frac{\partial}
{\partial{b_{k}^R}} \right) \right]A(s)=0 \end{eqnarray}

where, as usual, the beta functions are the derivatives of the corresponding
couplings with respect to $\log \mu$. The above equation is extremely useful
since it can be
 used to determine the dependence of the generating functions on $\mu$, and
finally, the dependence of all the couplings they generate on this scale. In
fact, this can be done in many different ways but, in order to make the chiral
limit more transparent, we will require the above equation to apply separately
to $A_0$ and $A_1$. Thus, from eq.41 and  eq.49    we can write

\begin{eqnarray} dG^{R}\left(s;\mu\right) \left(1-
\frac{sG^{R}\left(s;\mu\right)}{2\left(4\pi\right)^2F^2} T\left(s;\mu\right)
\right) +G^{R}\left(s;\mu\right) d\left(1-
\frac{sG^{R}\left(s;\mu\right)}{2\left(4\pi\right)^2F^2} T\left(s;\mu\right)
\right)=0 \end{eqnarray}

and integrating this equation we find:

\begin{eqnarray} G^{R}\left(s;\mu\right)= \frac{G^{R}\left(s;{\mu}_{0}\right)}
{1-\frac{sG^{R}\left(s;\mu_0\right)}{2\left(4\pi\right)^2F^2}
\log{\frac{\mu}{\mu_{0}}}} \end{eqnarray}

This equation describes the dependence on the $G^R(s;\mu)$ generating function
on the renormalization scale $\mu$ determined by the renormalization group
equations.

Now, applying eq.49 to $A_1$ we find that the piece

\begin{eqnarray} A^R(s;\mu)-B^R(s;\mu)\frac{s}{\left(4\pi\right)^2F^2}
\log{\frac{m^2}{\mu^2}} \end{eqnarray}

must be $\mu$ independent. For the sake of simplicity we introduce another
generating function $J^R(s;\mu)$ defined as:

\begin{eqnarray}
J^R(s;\mu)=A^R(s;\mu)-B^R(s;\mu)\frac{s}{\left(4\pi\right)^2F^2}
\log{\frac{m^2}{\mu^2}}+\frac{s}{\left(4\pi\right)^2F^2}
\log{\frac{m^2}{\mu^2}} \end{eqnarray}

note that in general the couplings  $j_k^R$ generated by $J^R$ can also depend
on $\log m^2$ but this is not the case of $j_1^R$.  In terms of $J^R$ eq.52
reads:

\begin{eqnarray} \frac{dJ^R(s;\mu)}{d\left(\log{\mu}\right)}=
\frac{-2s}{\left(4\pi\right)^2F^2} \end{eqnarray}

This equation can be integrated to give:

\begin{eqnarray} J^R\left(s;\mu\right)= J^R\left(s;{\mu}_{0}\right)
-\frac{2s}{\left(4\pi\right)^2F^2} \log { \frac{\mu} {{\mu}_0} } \end{eqnarray}

Now it is possible to expand both sides of the evolution equations eq.51 and
eq.55 for the $G^R$ and the $J^R$ generating functions in powers of $s/F^2$:
\begin{eqnarray} G^R\left(s;\mu\right)= \sum_{k=0}^{\infty}
g_{k}^{R}\left(\mu\right) \left(\frac{s}{F^2}\right)^{k} \\ \nonumber
J^R\left(s;\mu\right)= \sum_{k=0}^{\infty} j_{k}^{R}\left(\mu\right)
\left(\frac{s}{F^2}\right)^{k} \end{eqnarray}
 to find two infinite sets of evolution equations which explicitly give the
dependence of the  renormalized couplings generated by $G^R$ and $J^R$ on the
scale $\mu$. For example, for the lowest couplings we find: \begin{eqnarray}
g_{1}^{R}\left(\mu\right)= g_{1}^{R}\left(\mu_{0}\right)-
\frac{1}{16\pi^2}\log{\frac{\mu}{\mu_{0}}} \nonumber\\
j_{1}^{R}\left(\mu\right)= j_{1}^{R}\left(\mu_{0}\right)-
\frac{1}{8\pi^2}\log{\frac{\mu}{\mu_{0}}} \end{eqnarray} In terms of the $G^R$
and $J^R$ generating  functions the $A(s)$ amplitude can be written as:
\begin{eqnarray} A\left(s\right)= \frac{1}{NF^2} \frac{G^{R}(s;\mu)} {1-
\frac{sG^{R}(s;\mu)}{2\left(4\pi\right)^2F^2} T(s;\mu)} \left\{s-
\frac{m^2G^{R}(s;\mu)} {1- \frac{sG^{R}(s;\mu)}{2\left(4\pi\right)^2F^2}
T(s;\mu)} \left[J^{R}(s;\mu)
-\frac{s}{\left(4\pi\right)^2F^2}\log{\frac{m^2}{\mu^2}} \right] \right\}
\end{eqnarray} From this result it is not difficult to obtain the the amplitude
low energy behavior which is given by: \begin{eqnarray} A\left(s\right) &
\simeq & \frac{s}{NF^2} \left\{1-
\left(2g_1^R\left(\mu\right)+j_{1}^{R}\left(\mu\right)\right)\frac{m^2}{F^2}+
2\frac{m^2}{\left(4\pi\right)^2F^2} \log{\frac{m^2}{\mu^2}} \right\}
-\frac{m^2}{NF^2} \nonumber\\ & + &
O\left[\left(\frac{m^2}{F^2}\right)^2\right]+ O\left(\frac{s}{F^2}\right)^2
\end{eqnarray} This is a very useful and interesting result since we know that
at low energies $A(s)$ should go as $(s-m_{\pi}^2)/f_{\pi}^2$ where $m_{\pi}$
is the physical pion mass and $f_{\pi}$ is the pion decay constant. Therefore
the amplitude in eq.58 has the right low energy behavior provided we define:
\begin{eqnarray} f_{\pi}^2=NF_{eff}^2 =NF^2 \left\{ 1-\frac{m^2}{F^2} \left(
\frac{1}{8\pi^2} \log { \frac{m^2}{\mu^2} } -2g_1^R-j_{1}^{R} \right)
\right\}+O\left[\left(\frac{m^2}{F^2}\right)^2\right] \end{eqnarray} this
equation also provides the right  non trivial dependence of $f_{\pi}$ on the
logarithm of the pion mass which is well known from chiral perturbation theory
\cite{GassLeut} and in it $F_{eff}^2$ is not depending on the scale $\mu$ as
can be shown by using eq.57. These facts are
 therefore a good consistency check of our results. Moreover, from
  eq.28 we see that the pion mass should also be renormalized. This can be
done, for example, defining the renormalized mass as:   \begin{eqnarray}
m_{R}^2=m^2 \left\{1-\frac{m^2}{2\left(4\pi\right)^2F^2}
\left(N_{\epsilon}+1\right) \right\} \end{eqnarray}

so that the effective mass appearing in the effective action  can be written
as: \begin{eqnarray} m_{eff}^2=m_{R}^2+
\frac{m^4}{2\left(4\pi\right)^2F^2}\log{\frac{m^2}{\mu^2}} \end{eqnarray} in
good agreement with well known results from chiral perturbation theory
\cite{GassLeut}. However it is important to notice that our computation of the
effective action and the scattering amplitude for pions is only performed up to
the lowest order in $m^2/F^2$. Thus, as the renormalization of the mass
introduces an extra factor $m^2/F^2$
($m_{eff}^2=m^2+O\left(\frac{m^2}{F^2}\right)$),
 we do not have to renormalize that $m^2$ appearing in the amplitude in eq.30
since it would have produced extra terms proportional to $(m^2/F^2)^2$, and
the same applies to eq.58. Therefore we can consider the $m^2$ appearing in
eq.30 and eq.58 as the physical pion mass squared $m^2_{\pi}$.

\section{Partial Waves and Unitarity}

In order to study the unitarity properties of the amplitudes obtained in the
preceeding section we will perform the standard partial wave  expansion. First
we project on the isospin channels which for the model here considered  are be
defined as \cite{Dugan}:

\begin{eqnarray} T_{0} & = & NA\left(s\right)+A\left(t\right)+A\left(u\right)
\nonumber\\ T_{1} & = & A\left(t\right)-A\left(u\right) \nonumber\\ T_{2} & =
& A\left(t\right)+A\left(u\right) \end{eqnarray} The partial waves are then
defined as usual: \begin{eqnarray} a_{IJ}(s)= \frac{1}{64\pi} \int_{-1}^{1}
d\left(\cos{\theta}\right) T_{I}\left(s,\cos{\theta}\right)
P_{J}\left(\cos{\theta}\right) \end{eqnarray} As it is well known, the
requirement of unitarity constraints the
 possible behaviour of these partial waves. In particular, they should have a
cut  along the positive real axis from the threshold to infinity. The physical
amplitudes are obtained from the values right on the cut, and, in addition, the
condition of elastic unitarity: \begin{equation} Im  a_{IJ}  =   \sigma
|a_{IJ}|^2 \end{equation} (where $ \sigma  =  \sqrt{1-4m^2/E^2}$) must be
fulfilled in the physical   region $s= E^2+i\epsilon$ and  $E^2  >  4m^2$
where $E$ is the center of mass energy. This equation is exact for energies
below the next four pion threshold but even beyond that point it is
approximately valid.

In the following we will study how well this relation is satisfied in the
leading large $N$ pion scattering amplitudes. From eq.63 it is obvious that
the most important isospin channel at the leading order in the large $N$
expansion is $I=0$ which is of order $1$ in the $1/N$ expansion.  From eq.58,
eq.63  and eq.64  and for the $a_{00}$ channel it is immediate to find:
\begin{eqnarray} a_{00}\left(s\right)= \frac{1}{32 \pi F^2} \left\{
\frac{sG_{R}(s;\mu)} {1-
\frac{sG_{R}\left(s;\mu\right)}{2\left(4\pi\right)^2F^2}
\left(\tilde{T}+i\pi\sigma\right) }- \frac{m^2 \left( J_{R}\left(s;\mu\right)
-\frac{s}{\left(4\pi\right)^2F^2} \log{\frac{m^2}{\mu^2}} \right)} {\left[1-
\frac{sG_{R}\left(s;\mu\right)}{2\left(4\pi\right)^2F^2}
\left(\tilde{T}+i\pi\sigma\right) \right]^2} \right\} \end{eqnarray} where we
have defined $\tilde{T}$ as \begin{eqnarray} T\left(E^2+i\epsilon;\mu^2\right)
& = &  \sigma\log{|\frac{\sigma-1}{\sigma+1}|}
+i\pi\sigma-\log{\frac{m^2}{\mu^2}} \nonumber\\ & \equiv &
\tilde{T}+i\pi\sigma \end{eqnarray}

and thus we are only considering explicitly
 the physical values of $s$. However, it is easy to see that the $a_{00}(s)$
partial wave amplitude, when $s$ is considered a complex variable, has the
proper right unitary cut mentioned above. The partial wave above can be now
naturally decomposed as: \begin{equation} a_{00}  \equiv
a_{00}^{\left(0\right)}+a_{00}^{\left(1\right)} \end{equation} where
$a_{00}^{\left(1\right)}$ corresponds to the part of the partial wave that is
proportional to $m^2$ and  $a_{00}^{\left(0\right)}$ is the part which is not.
In terms of
 $a_{00}^{\left(0\right)}$ and $a_{00}^{\left(1\right)}$   the unitarity
condition in eq.65   reads: \begin{eqnarray} Im a_{00}^{\left(0\right)} & = &
\sigma |a_{00}^{\left(0\right)}|^2 \nonumber\\ Im a_{00}^{\left(1\right)} & =
&  2\sigma Re a_{00}^{\left(0\right)}a_{00}^{\left(1\right)\ast} \end{eqnarray}
provided one neglects terms of order $(m^2/F^2)^2$. After a bit of algebra it
is not difficult to show that these equations are indeed satisfied by
$a_{00}^{\left(0\right)}$ and $a_{00}^{\left(1\right)}$ as obtained from eq.65
and eq.68. Therefore, we finally can write for the partial wave $a_{00}$ from
the amplitude in eq.66 :  \begin{eqnarray} Im a_{00}= \sigma |a_{00}|^2+
O\left(\frac{1}{N}\right)+ O\left[\left(\frac{m^2}{F^2}\right)^2\right]
\end{eqnarray} As usual  the  right
 cut produces two sheets for the $a_{00}$ function. Depending on the actual
form
 of the generating functions, eventually some poles could appear in different
 places. If the poles appear in the first sheet (the physical sheet) they are
not acceptable and must be understood as artifacts (ghosts) of the
approximation. However, when the poles appear  on the second sheet (the
unphysical sheet) they are welcome and have  a natural interpretation as
resonances. This is for example the case of the LSM (Linear Sigma Model) that
will be studied in detail in the next  section.
\section{The Linear $\sigma$-model}
In this section we consider the case of the LSM which is
defined by the lagrangian
\begin{eqnarray}
 {\cal L}=\frac{1}{2}
{\partial}_{\mu} {\phi}^{T} {\partial}^{\mu}{\phi}- V\left(\phi\right)+ {\cal
L}_{SB}
\end{eqnarray}
 where $\phi$ is an $N+1$ vector with components $\pi_1,
\pi_2...,\pi_N,\pi_{N+1}$ and the potential is given by: \begin{eqnarray}
V\left(\phi\right)= -{\mu}^2|\phi|^2+\lambda|{\phi}^2|^2 \end{eqnarray} The
lagrangian would be $O(N+1)$ invariant  were it not for the last symmetry
breaking  piece that it is
 only $O(N)$ invariant. This piece is given by: \begin{eqnarray} {\cal L}_{SB}
& = &  \sqrt{N}F\tilde{m}^2\sigma \end{eqnarray} where  \begin{eqnarray}
\sigma & = & {\pi}_{N+1}= \sqrt{NF^2}+H \end{eqnarray} In terms of the field
$H$ the lagrangian can be written as: \begin{eqnarray} {\cal L} & = &
\frac{1}{2} {\partial}_{\mu}{\pi^a} {\partial}^{\mu}{\pi^a}+\frac{1}{2}
{\partial}_{\mu}H {\partial}^{\mu}H+ \lambda N^2F^4+ \tilde{m}^2NF^2
\nonumber\\ & - & 4\lambda NF^2H^2+ H\sqrt{NF^2}\tilde{m}^2-
\lambda\left({\pi}^2+H^2\right)^2 \nonumber\\ & - &
4\lambda\sqrt{NF^2}H{\pi}^2- 4\lambda\sqrt{NF^2}H^3 \end{eqnarray} Notice that
there is a linear term in the $H$ field. In order to eliminate it, we
introduce a shifted $h$ field: \begin{eqnarray} H= H_{0}+h \end{eqnarray} with
constant $H_0$. From the lagrangian written using this new field $h$, and in
particular, from the term proportional to $\pi^2$  we find the pion mass to be:
\begin{eqnarray} m^2=4\lambda H_{0}^{2}+8\lambda H_{0}\sqrt{NF^2}
\end{eqnarray} The piece of the lagrangian proportional to $h$ can be written
in terms of this
 pion mass. Then, the condition for this piece to vanish become:
\begin{eqnarray} H_{0}=-\sqrt{NF^2}+ \sqrt{NF^2+ \frac{m^2}{2\lambda}}>0
\end{eqnarray} So that the lagrangian has the final form: \begin{eqnarray}
{\cal L} & = & \frac{1}{2} {\partial}_{\mu}{\pi^a} {\partial}^{\mu}{\pi^a}
-\frac{1}{2}m^2{\pi^2} +\frac{1}{2}{\partial}_{\mu}h {\partial}^{\mu}h-
\frac{1}{2}m_{h}^{2}h^2 \nonumber\\ & - & \lambda\left({\pi}^2+h^2\right)^2
-4\lambda h\left({\pi}^2+h^2\right) \sqrt{NF^2+\frac{m^2}{4\lambda}}
\end{eqnarray} where \begin{eqnarray} m_{h}^2 & = & M^2+3m^2 \nonumber\\ M^2 &
= & 8\lambda NF^2 \end{eqnarray} Now we are able to obtain the corresponding
Feynman rules. The elastic scattering amplitude
 can be computed in many different ways in the large $N$ limit.  Probably the
easier one is the following: First one compute the diagrams in fig.7a, then
one iterates the result as shown in fig.7b and, finally, one expands in powers
of $m^2/F^2$ and retains the zero and the first order. The final result is
given by: \begin{eqnarray} A(s)=\frac{1}{NF^2} \left\{ \frac{sG_{M}(s)} {1+
\frac{sG_{M}(s) I(s)}{2F^2} }- \frac{m^2G_{M}^2(s) \left( 1+2 \frac{s}{M^2} +
\bigtriangleup \frac{s}{F^2} \right)} {\left(1+ \frac{sG_{M}(s) I(s)}{2F^2}
\right)^2 } \right\} \end{eqnarray} where: \begin{eqnarray}
G_{M}\left(s\right)= \frac{1}{1-\frac{s}{M^2}} \end{eqnarray} Notice that we
can consider this result is a particular case of the general one given in
eq.36 with the elections: \begin{eqnarray} G\left(s\right) & = &
G_{M}\left(s\right) \nonumber\\ H\left(s\right) & = &
G_{M}^2\left(s\right)\left(1+2\frac{s}{M^2}\right) \nonumber\\ E\left(s\right)
& = & G_{M}^2\left(s\right) \end{eqnarray} The possibility of writing the LSM
amplitude as a particular
 case of the NLSM (in the large $N$ limit in both cases) was expected form our
previous discussion but, as far as both computations were done in a completely
different and independent way, it turns to be a highly non trivial check of
our results.
 Now we can proceed with the renormalization procedure as we did in sec.3 and
sec.4. The relevant issue here is that the LSM is renormalizable in the
standard way, i.e. only a finite number  parameters have to be renormalized.
This means in  the large $N$ limit here considered  that only the constant $M$
will be renormalized. In particular, we find:  \begin{eqnarray}
G_{M}^{R}\left(s;\mu\right)= \frac{1}{1-\frac{s}{M_{R}^2\left(s;\mu\right)}}
\end{eqnarray} where \begin{eqnarray} \frac{1}{M_{R}^2}= \frac{1}{M^2}+
\frac{1}{2\left(4\pi\right)^2F^2} \left(N_{\epsilon}+2\right) \end{eqnarray}
so that the dependence of the $M^2_R$ with the scale is given by
\begin{eqnarray} M_{R}^2(\mu)= \frac{M_{R}^2(\mu _ 0)} {1- \frac{M_{R}^2(\mu
_0)} {2 (4 \pi)^2F^2} \log{ \frac{\mu ^2}{\mu _0^2}} } \end{eqnarray} The
other needed renormalized generating function is: \begin{eqnarray}
J^{R}\left(s;\mu\right)= 1+2\frac{s}{M_{R}^2(\mu)} \end{eqnarray} Now, using
eq.86   we  find: \begin{eqnarray}
\frac{dJ_{R}\left(s;\mu\right)}{d\left(\log{\mu}\right)}=
\frac{-2s}{\left(4\pi\right)^2F^2} \end{eqnarray} which is consistent with the
general $\mu$ dependence
 for the generating functions found in eq.51 and eq.55.  Finally, the
renormalized amplitude for the LSM is found to be: \begin{eqnarray}
A\left(s\right)= \frac{1}{NF^2} \left\{ \frac{sG^{R}\left(s;\mu\right)} {1-
\frac{sG^{R}\left(s;\mu\right)}{2\left(4\pi\right)^2F^2} T\left(s;\mu\right) }-
\frac{m^2 {G^{R}}^2\left(s;\mu\right)
\left(1+2\frac{s}{M_{R}^2\left(\mu\right)}-\frac{s}
{2\left(4\pi\right)^2F^2}\log{\frac{m^2}{\mu^2}} \right) } {\left(1-
\frac{sG^{R}\left(s;\mu\right)}{2\left(4\pi\right)^2F^2} T\left(s;\mu\right)
\right)^2 } \right\} \end{eqnarray} In the chiral limit, this amplitude
reduces to that found in \cite{Italians}. The $I=J=0$ partial wave amplitude
has  a pole in the second Riemann
 sheet that corresponds to a physical
 scalar resonance. This resonance has the interesting saturation property,
which means that its mass  remains bounded when $M_R$ goes to infinity (see
also \cite{Dobado} for a general discussion). This fact is  probably related
with the expected triviality of the LSM.

In any case, the LSM provides a very nice example of how the general formalism
developed for the computation of the large $N$ limit of $\chi PT$ works
outside of the chiral limit.

\section{Fitting the pion scattering}

Apart from the LSM, the most interesting case to apply our results is the pion
scattering. However, in order to use eq.58 to fit the experimental pion
scattering
 data, we have to face the problem of dealing with an infinite number of
parameters, i.e.
 the scale $\mu$ and the values of the renormalized coupling constants
$g_k^R(\mu)$ and $j_k^R(\mu)$  at that scale. However one can easily solve
this problem  just
 considering only particular cases where all the coupling constants but a
finite set $g_1^R$, $g_2^R \dots g_r^R$ and  $j_1^R$, $j_2^R \dots j_s^R$
vanish at some scale $\mu$. These models are just defined by a finite number
of parameters ($\mu$
 and the $r+s$ coupling constants renormalized at this scale) and therefore can
be used to fit the experimental data. In particular, one can consider the
extreme case of having all the renormalized couplings equal to zero at some
scale $\mu$, i.e. $g_k^R(\mu)=j_k^R(\mu)=0$ for all $k>0$ (the minimal model).
The  model thus obtained only has one parameter (the $\mu$ scale) and it can
be considered in some sense as the non-linear sigma model renormalized at the
leading order of the $1/N$ expansion out of the chiral limit.

In fig.8 we show the result of our fit of the $I=J= 0$
 phase shift for elastic pion scattering done with the above
 defined minimal model. The only parameter in this specially
 simple case is the  scale $\mu$. The fit in fig.8 corresponds to a value of
$\mu\simeq 775 MeV$ and as it can be seen it describe perfectly well the data
from the threshold to $700 Mev$. It is also interesting to remark that the
fitted value for $\mu$ is the same that was found in \cite{DoLoMo}. However,
as we are including here the effect of the pion mass, the fit is much better
in the threshold region.

For other
 channels like $I=J=1$ or $I=2, J=0$ the prediction of the leading order of
the  large $N$ expansion is that they are supressed. This could seem quite a
poor prediction but it is not. Looking at the experimental data we can realize
that the phase shifts in these channels grow very slowly with  energy, for
instance, at $500 MeV$ of center of mass energy we have $\delta_{11} \simeq 5
$ degrees
  and $\delta_{20}\simeq -7 $ degrees whereas $\delta_{00}\simeq 38
 $ degrees at the same energy. In any case one can use the results obtained in
the previous section considering eq.58, eq.63 and eq.64 with the $\mu$  value
fitted for the $a_{00}$ channels. The results are shown in fig.8 and fig.9.
For the $I=2, J=0$ channel we see that our fit (wich is in fact a prediction
since we only have one  parameter that had already been fixed in the $I=J=0$
channel) is quite good. However this is not the case when $I=J=1$. Basically
this channel can be described as a non interacting channel at low energies but
a strong interacting channel at higher energies due to the appearence of the
$\rho$ resonence. The approach considered here seems to describe well the
non-interacting low-energy region but it fails in the resonant region. One is
then tempted to  make an interpretation  of the scale $\mu$ as some kind of
cutoff signaling the range of applicability of our approach and, in fact,
this interpretation  had already been discussed in \cite{largen}. From this
point of view, it is quite interesting to realize  how close the $\mu$
 fitted value $(\mu\simeq 775 MeV)$ is to the $\rho$ mass. In some way, by
fitting the $I=J=0$ channel, the model is telling us where new physics can
appear in other channels like the $I=J=1$ (the $\rho$ resonance) and thus
setting the limits of applicability of the model. However, we would like to
stress that our renormalization method is
 completely
 consistent and our results are formally valid at any energy independently of
the goodness of the fits we can obtain with them.

Finally, for comparison we have also included in fig.8 and fig.9 the results
obtained with standard $\chi PT$ to one loop with the parameters proposed in
\cite{GassMeiss}. It  is important to note that, in spite of  using many more
couplings, it does not a better job than the approach considered here. The
description of the $I=J=0$ channel is good in both cases, the $I=2, J=0$ is
much better in the large $N$ limit and the $I=J=1$ is good at low energies
(where the channel is not interacting) but it fails to describe the resonance
region too.

\section{Conclusions}

In this work we have study how the large $N$ approximation
 ($N$ being the number of Goldstone bosons) can be applied to $\chi PT$ in
order to improve its unitarity behavior and enlarge its energy applicability
region.  In particular we have computed the renormalized effective action at
the leading order
 including the effects due to the pion mass up to order $m^2_{\pi}/f_{\pi}^2$.
The amplitude obtained from this action has the appropriate low energy limit
and provides the right dependence of the the pion mass and the pion decay
constant on the chiral logs. We use this action to reproduce the LSM and to
fit the $\pi-\pi$ scattering with just one parameter. Within this approach the
fit  is competitive with the standard one-loop fit both failing in the
description of the $\rho$ resonance. However, it can improve when next to
leading ($1/N^2$) corrections are included. Note that due to the structure of
the large $N$
 amplitudes it is possible to find poles  in the second Riemann sheet that
could reproduce resonances as it happens in the large $N$ description of the
LSM. At present, work is also in progress in that direction \cite{DoMo}.

In conclusion, we consider that the great sucsess of $\chi PT$ can be enlarged
with the use of well defined non-perturbative techniques as the large $N$
expansion. The
 improvement can be obtained in several directions like unitarity, energy
range of application and finally, in order to make predictions, avoiding the
inflation in the number of free parameters.

\section{Acknowledgments} This work has been supported in part by the
Ministerio de Educaci\'on y Ciencia (Spain) (CICYT AEN90-0034) and COLCIENCIAS
(Colombia). A. D. thaks to the CERN Theory Division, where the final part of
this work was written, for its kind hospitality.

\newpage

\begin{center} {\large \bf Figure Captions} \end{center} {\bf Figure 1:} One
typical Green function obtained by attaching pions
 (continous line) to one $B$ (wavyline) and $\Phi$ (dashed line) Green
function. Note
 the possibility of having pion loops in the attaching point.

{\bf Figure 2:} Example of $\Phi$ Green (black dot) in terms of the reduced
(i.e. including only pion loops) $\Phi$ and $B$ Green
 functions (dashed dot).

{\bf Figure 3:} Example of leading diagramns contributting to the reduced $B$
Green functions in the Large $N$ limit.

{\bf Figure 4:} Relation between the $B$ Green functions and the
 reduced $B$ Green functions at leading order in the large $N$ approximation.

{\bf Figure 5:} a) Diagrams contributing to the reduced one $\Phi$ Green
function b) Diagrams contributing to the reduced two $B$ Green function.
Similar diagrams
 contribute to the two $\Phi$ and the $B\Phi$ reduced Green functions.

{\bf Figure 6:} Diagrams contributing to the the pion scattering to leading
order in the large $N$ approximation up to order $m^2/F^2$. The large black
dot represents the pion coupling proportional to $m^2$

{\bf Figure 7:} Diagrams appearing in the computation of the pion scattering
 amplitude in the LSM described in the text. The continuous lines represent
pions and the dashed ones the $h$ field.

{\bf	Figure 8:}  Phase shift for $\pi \pi$ scattering. The results coming from
the large $N$ limit out of the chiral limit are drawn with a    dashed line
both for the $(0,0)$ and the $(2,0)$ channels. The continuous lines were taken
from \cite{piK} and they are the  $(0,0)$ and $(2,0)$ phase shifts as obtained
from the standard one-loop $\chi PT$ with the parameters given in
\cite{GassMeiss}. The experimental data corresponds to: \cite{Rosselet}
$\bigtriangleup$, \cite{Manner} $\bigcirc$, \cite{Srinivasan} $\Box$,
\cite{David} $\diamondsuit$, \cite{Esta} $\bigtriangledown$, \cite{Hoogland}
$\star$, \cite{Losty} $\times$, and \cite{Hoogland2} $\bullet$.

{\bf	Figure 9:}  $(1,1)$ Phase shift for $\pi \pi$ scattering. The dashed line
corresponds to our fit using the large $N$ limit. The continuous line
\cite{piK} is the result coming from one-loop $\chi PT$ with the parameters
proposed in \cite{GassMeiss}. The experimental
 data comes from: \cite{Esta} $\bigcirc$, \cite{Proto} $\bigtriangleup$.

\end{document}